\documentclass[useAMS, fleqn, usenatbib]{mn2e}
\usepackage{amsmath}%
\usepackage{amsfonts}%
\usepackage{amssymb}%
\usepackage{graphicx}
\usepackage{color}

\newcommand\dderiv{{\mathrm{d}}}

\title[Probing Large-Angle Correlations with $C^{T\varphi}(\theta)$]{Probing Large-Angle Correlations with the Microwave Background Temperature and Lensing Cross Correlation}

\author[Yoho A., Copi C. J, Starkman G. D., Kosowsky A.]{A.~Yoho,$^{1,2}$ C.~J.~Copi,$^{1}$ 
G.~D.~Starkman,$^{1,2}$ A.~Kosowsky,$^{3}$\\
$^1$CERCA/ISO, Department of Physics, Case Western Reserve University,
10900 Euclid Avenue, Cleveland, OH 44106-7079, USA\\
$^{2}$ CERN, CH-1211 Geneva 23, Switzerland\\
$^{3}$ Department of Physics and Astronomy, University of Pittsburgh, Pittsburgh, PA 15208 USA}

\begin{document}
\maketitle

\begin{abstract}
A lack of correlations in the microwave background temperature between sky directions
separated by angles larger than $60\degr$ has recently been confirmed by data from
the {\it Planck} satellite. This feature arises as a random occurrence within
the standard $\Lambda$CDM cosmological model less than $0.3$ per cent of the time, but so far no
other compelling theory to explain this observation has been proposed. Here we investigate
the theoretical cross-correlation function between microwave background temperature and the
gravitational lensing potential of the microwave background, which in contrast to the temperature
correlation function depends strongly on  gravitational potential fluctuations interior to our Hubble
volume.  For standard $\Lambda$CDM cosmology, we generate random sky realizations
of the microwave temperature and gravitational lensing,
subject to the constraint that the temperature correlation function matches observations, and
compare with random skies lacking this constraint. The distribution of large-angle temperature-lensing
correlation functions in these two cases is different, and the two cases can be clearly
distinguished in around $40$ per cent of model realizations. We present an {\it a priori} procedure for 
using   similar large-angle correlations between other types of data,
to determine whether the lack of large-angle correlations is a 
statistical fluke or points to a shortcoming of the standard cosmological model.

\end{abstract}

%%%%%%%%%%%%%%%%%%%%%%%%
\section{Introduction}
%%%%%%%%%%%%%%%%%%%%%%%%

Observations of the Cosmic Microwave Background (CMB) have provided cosmologists
with a wealth of information about our early Universe. 
On its own, and especially in concert with data from complementary probes, CMB observations have
lead to  increasingly tight constraints on the inferred values of  cosmological parameters, 
and  have allowed us to to distinguish among various models of our Universe.  This has
resulted in a  standard cosmological model: inflationary flat Lambda Cold Dark Matter  ($\Lambda$CDM).

Despite its great successes, $\Lambda$CDM has had difficulty explaining certain features in the CMB that were 
initially characterized by the {\it Cosmic Background Explorer's Differential Microwave Radiometer (COBE-DMR)} \citep{Bennett:1996ce, Hinshaw:1996ut}
or using {\it Wilkinson Microwave Anisotropy Probe (WMAP)} observations \citep{Spergel:2003cb,Chang:2013vla, Copi:2006tu, Copi:2008hw, Copi:2010na} 
and recently confirmed \citep{Ade:2013nlj, Copi:2013cya} 
with the first release of  temperature data from the {\it Planck} satellite. 
These features are predominantly in the large-angle,  or low  ($\ell\lesssim 30$) multipole, regime. The anomalies include 
an Ecliptic north-south hemispherical asymmetry \citep{Eriksen:2007pc}, 
the alignment of the quadrupole and octopole patterns with one another \citep{de OliveiraCosta:2003pu,Copi2004,Land:2005ad, CHSSprep}, with the Ecliptic \citep{Schwarz:2004gk}
and with the cosmological dipole \citep{Copi:2005ff},  and low variance across the sky \citep{Hou:2009au}. A full list 
as well as comparison to {\it WMAP} observations can be found in \citep{Ade:2013nlj}.

Another large-angle anomaly, namely the lack of correlation on the CMB sky at angles larger than about $60\degr$, 
currently has no compelling explanation. The importance of this feature has been 
outlined in papers over the last several years~\citep{Schwarz:2004gk}. Thus far, this  anomaly has been observed only in the
temperature auto correlation function.  There has been some suggestion that it is merely a statistical fluke (which we call
the fluke hypothesis, or the null hypothesis because it is just standard cosmology informed by the experimental data)
especially since it was only noted and characterized {\it a posteriori}.   Moreover, it will be difficult to improve on the
current statistical significance purely through further observation of the temperature correlations because of cosmic variance and because the measurements are statistics limited. 
%This limits our ability to make robust statistical statements about the anomalously
%low power -- with only one realization of our universe we are unable to make definitive statements about
%a particular feature being a result of interesting primordial physics or just a statistical fluke. 

The use of temperature data exclusively is due to to a lack of high signal-to-noise full-sky maps of 
other cosmological quantities, such as polarization and lensing potential. With the highly anticipated
upcoming release of the {\it Planck} polarization map as well as several upcoming lensing experiments, 
cosmologists will have information on large-angle correlation functions beyond just microwave temperature data. Recently 
the ability for a temperature-$Q$ polarization cross correlation to test the fluke hypothesis
was investigated~\citep{Copi:2013zja}. The results of that work showed that there was a possibility for $TQ$ correlations to
rule out the null hypothesis, but that it would not necessarily be a definitive test. In this work we will
investigate the possibility for a cross correlation between temperature and CMB lensing potential to provide
a test of the statistical fluke hypothesis as well as a consistency check for observations.
 We provide predictions for the distribution of a 
standard statistic used in two-point correlation function analysis, $S_{1/2}$, as well as define an optimal,
{\it a priori} statistic for use with temperature-lensing cross correlations.

While in this paper we present results comparing two particular models -- 
unconstrained and constrained $\Lambda$CDM -- we use a generic prescription for optimization and 
for determining the viablity of an $S_{1/2}$-like
statistic. This analysis can be repeated for any set of models as long as one knows
how to produce realizations within that framework. This reason in particular was a driving force in the choice of 
unconstrained $\Lambda$CDM for our comparison model, as generating realizations is straightforward.
 With the particular choice of $T\varphi$ correlations, the discriminating power of the $S^{T\varphi}$ statistic
will depend on how a particular model suppresses the $\Theta_{\rmn{ISW}}\varphi$ term in the correlation, as this is the piece that dominates the signal, unlike $TT$ correlations. However, this analysis can {\it always}
be carried out such that the $S_{1/2}$-like statistic can be optimized {\it a priori} for correlations between
cosmological data sets.

The paper is organized as follows: in Sec.~\ref{background} we will give the theoretical background for CMB two-point
correlation function analysis, in Sec.~\ref{constrained} we will describe how we generate constrained $\Lambda$CDM 
realizations, in Sec.~\ref{s12} we will outline our general prescription for calculating 
statistics, in Sec.~\ref{results} we will present our results for the distributions of the statistic for two models (constrained
and unconstrained $\Lambda$CDM), and finally in Sec.~\ref{conclusions} we will 
summarize and state our conclusions.

%%%%%%%%%%%%%%%%%%%%%%%%
\section{Background}\label{background}
%%%%%%%%%%%%%%%%%%%%%%%%

The two-point angular correlation function for the CMB temperature is calculated
by taking an ensemble average of the temperature fluctuations in different directions:
\begin{equation}
C^{TT}(\theta)=\langle\Theta(\boldsymbol{\hat{n}_1})\Theta(\boldsymbol{ \hat{n}_2})\rangle   \quad  \mbox{with} \quad \boldsymbol{\hat{n}}_1\cdot\boldsymbol{\hat{n}_2}=\cos\theta.
\end{equation}
Since we are not able to compute the ensemble average in practice, we instead calculate ${\cal C}^{TT}(\theta)$, a sky average over the angular separation. In general, any $C(\theta)$  can be expanded in a Legendre series, which we write as
\begin{equation}\label{theta_ell}
C(\theta)=\sum_{\ell}\frac{\ell(\ell+1)}{4\upi}\,C_{\ell}P_{\ell}(\cos\theta),
\end{equation}
where the $C_{\ell}$ on the right-hand side of Eq.~\ref{theta_ell} are the measured power spectrum values.
On a full sky, the coefficients $C_{\ell}$ obtained from the estimator
\begin{equation}\label{powerspectrum}
C_{\ell} \equiv \frac{1}{2\ell+1}\sum_{m=-\ell}^\ell \vert a_{\ell m}\vert^2 ,
\end{equation}
where the $a_{\ell m}$ are the usual spherial harmonic coefficients of a map, 
are identical to the $C_{\ell}$ in Eq.~\ref{theta_ell}.
Computationally it is more efficient to use the relations~(\ref{theta_ell}) and (\ref{powerspectrum})
than to directly correlate pairs of pixels on the observed CMB sky.

The two-point function from the {\it COBE-DMR}'s fourth-year data release \citep{Bennett:1996ce}
highlighted a lack of temperature auto-correlations for angular separations larger than 60 degrees. 
The {\it WMAP} first-year data release \citep{Spergel:2003cb} first quantified this feature using an ({\it a posteriori}) statistic 
that neatly captured the simple observation that $C(\theta)$ nearly vanished at large angles
\begin{equation}\label{WMAPs12eq}
S_{1/2}^{TT}\equiv\int_{1/2}^{-1}\mbox{d}(\cos\theta)[C^{TT}(\theta)]^2.
\end{equation}
We generalize this to
\begin{equation}\label{s12eq}
S_{1/2}^{XY}\equiv\int_{1/2}^{-1}\mbox{d}(\cos\theta)[C^{XY}(\theta)]^2,
\end{equation}
where $X,Y$ can be any combination of cosmological quantities for which $C^{XY}(\theta)$ can be calculated.
The {\it WMAP} team calculated $S_{1/2}^{TT}$ on the most reliable part of the sky -- that outside the galactic plane --
and found it to be $1152 \;\umu \rmn{K}^4$, much smaller than the expected $\Lambda$CDM value of $\sim50\; 000 \;\umu \rmn{K}^4$. 
They remarked: `For our 
$\Lambda$CDM Markov chains$\dots$only  0.15\% of the simulations have lower values of S.'

The simplest (and perhaps leading) explanation for this anomaly is that it is just a statistical fluke within
completely canonical $\Lambda$CDM.  
In this paper, our goal is to define  {\it a priori} a statistic that can be used with future data to test the fluke hypothesis.
We therefore need to find an independent cross correlation that would respond to the same statistical
fluctuations  as the temperature two-point function.  To this end we focus on the lensing of the CMB, and construct
the cross correlation, $C^{T\varphi}(\theta)$, between the CMB temperature $T$ and the CMB lensing potential $\varphi$.
Since both derive from the gravitational potential $\Phi$, we expect that if statistical fluctuations generated by 
primordial physics in $\Phi$ caused the lack of large-angle temperature auto-correlation, 
then those statistical fluctuations would also imprint themselves in a new, distinct way on $C^{T\varphi}(\theta)$.

Since we know that at large
angles the CMB signal is dominated by the Sachs-Wolfe (SW) and Integrated Sachs-Wolfe (ISW) contributions, we 
can write 
\begin{equation}
\Theta(\boldsymbol{  \hat{n}})=\Theta_{\rmn{SW}}(\boldsymbol{  \hat{n}})+\Theta_{\rmn{ISW}}(\boldsymbol{  \hat{n}})
\end{equation}
 and can
describe $C^{TT}(\theta)$ as a correlation of these two terms only. The 
SW and ISW pieces of the primordial temperature fluctions
in terms of the metric potential, $\Phi$, are
\begin{equation}
\Theta_{\rmn{SW}}(\boldsymbol{  \hat{n}})=-\frac{1}{3}\Phi(\chi\boldsymbol{  \hat{n}}, \chi)
\end{equation}
and
\begin{equation}
\Theta_{\rmn{ISW}}(\boldsymbol{  \hat{n}})=-2\int_{0}^{\chi_{\ast}} \mbox{d}\chi\, \dot{\Phi}(\chi\boldsymbol{\hat{n}}, \chi),
\end{equation}
which allows us to write $C^{TT}(\theta)$ in terms of correlations of $\Phi$ as
%\begin{eqnarray}
%& & C^{TT}(\theta) = \frac{1}{9}\langle\Phi(\chi_{\ast}\boldsymbol{  \hat{n}_{1}},\chi_{\ast})\,\Phi(\chi_{\ast}\boldsymbol{  \hat{n}_2},%\chi_{\ast})\rangle \\
%& & + \frac{2}{3}\int \mbox{d}\chi_1\langle\Phi(\chi_1\boldsymbol{  \hat{n}_{1}},\chi_1)\,\Phi(\chi_{\ast}\boldsymbol{  \hat{n}_2},\chi_{\ast})\rangle %\\ & & + 4
%\int \mbox{d}\chi_1 \mbox{d}\chi_2 \langle\Phi(\chi_{1}\boldsymbol{  \hat{n}_{1}},\chi_1)\,\Phi(\chi_{2}\boldsymbol{  \hat{n}_2},\chi_2)\rangle .
%\end{eqnarray}
%\begin{multline}
% C^{TT}(\theta) = \frac{1}{9}\langle\Phi(\chi_{\ast}\boldsymbol{  \hat{n}_{1}},\chi_{\ast})\,\Phi(\chi_{\ast}\boldsymbol{  \hat{n}_2},\chi_{\ast})\rangle \\
% + \frac{2}{3}\int \mbox{d}\chi_1\langle\dot{\Phi}(\chi_1\boldsymbol{  \hat{n}_{1}},\chi_1)\,\Phi(\chi_{\ast}\boldsymbol{  \hat{n}_2},\chi_{\ast})\rangle \\  + 4
%\int \mbox{d}\chi_1 \mbox{d}\chi_2 \langle\dot{\Phi}(\chi_{1}\boldsymbol{  \hat{n}_{1}},\chi_1)\,\dot{\Phi}(\chi_{2}\boldsymbol{  \hat{n}_2},\chi_2)\rangle .
%\end{multline}
\begin{eqnarray}\label{eq:Ctt}
 C^{TT}(\theta) & = & \frac{1}{9}\langle\Phi(\chi_{\ast}\boldsymbol{  \hat{n}_{1}},\chi_{\ast})\,\Phi(\chi_{\ast}\boldsymbol{  \hat{n}_2},\chi_{\ast})\rangle \nonumber \\
& + & \frac{2}{3}\int \mbox{d}\chi_1\langle\dot{\Phi}(\chi_1\boldsymbol{  \hat{n}_{1}},\chi_1)\,\Phi(\chi_{\ast}\boldsymbol{  \hat{n}_2},\chi_{\ast})\rangle \nonumber \\  
& + & 4 \int \mbox{d}\chi_1 \mbox{d}\chi_2\langle\dot{\Phi}(\chi_{1}\boldsymbol{  \hat{n}_{1}},\chi_1)\,\dot{\Phi}(\chi_{2}\boldsymbol{  \hat{n}_2},\chi_2)\rangle .
\end{eqnarray}

Similarly, we can write the CMB lensing potential $\varphi$ in terms of the metric potential,
\begin{equation}
\label{eqn:lensing_potential}
\varphi(\boldsymbol{\hat{n}}) = 2\int_{0}^{\chi_{\ast}} \mbox{d}\chi\, \frac{\chi_{\ast}-\chi}{\chi_{*}\chi} \Phi(\chi\boldsymbol{\hat{n}}, \chi) ,
\end{equation}
which allows us to write an expression for the two-point function of the CMB temperature and lensing field:
\begin{eqnarray}
\label{eqn:Ctphi}
C^{T\varphi}(\theta)& = & \langle\varphi(\boldsymbol{  \hat{n}_1})\, \Theta_{\rmn{SW}}(\boldsymbol{  \hat{n}_2})\rangle + \langle\varphi(\boldsymbol{  \hat{n}_{1}})\,\Theta_{\rmn{ISW}}(\boldsymbol{  \hat{n}_2})\rangle \nonumber \\
& = & -\frac{2}{3}\int \mbox{d}\chi_1\langle\Phi(\chi_1\boldsymbol{  \hat{n}_{1}},\chi_1)\,\Phi(\chi_{\ast}\boldsymbol{  \hat{n}_2},\chi_{\ast})\rangle \nonumber \\ && - 4
\int \mbox{d}\chi_1 \mbox{d}\chi_2\langle \Phi(\chi_{1}\boldsymbol{  \hat{n}_{1}},\chi_1)\,\dot{\Phi}(\chi_{2}\boldsymbol{  \hat{n}_2},\chi_2)\rangle,
\end{eqnarray}
where computations to find $C^{T\varphi}(\theta)$ from data will use sky averages rather than ensemble averages.

From (\ref{eqn:Ctphi}), we see that the lensing field also accesses the information encoded in $\Phi$.  
Thus, if the anomalous absence of large-angle correlations in $C^{TT}(\theta)$ is due to statistical fluctuations, 
then it should be present in a predictable way in the $C^{T\varphi}(\theta)$ cross correlation. It should be noted, however,
that $C^{T\varphi}(\theta)$ traces the same physics in a different way -- the terms Eq.~\ref{eqn:Ctphi} do not exactly match
the SW and ISW terms in Eq.~\ref{eq:Ctt}. Furthermore, $C^{T\varphi}(\theta)$ is dominated by the ISW-$\varphi$ term, 
meaning the behavior of the two-point cross correlation is dominated by physics on the interior of our Hubble volume, whereas
$C^{TT}(\theta)$ has its largest contribution from the SW-SW term and is therefore dominated by physics at the last scattering
surface. This makes $C^{T\varphi}(\theta)$ a complementary probe into the nature of the lack of correlation in 
$C^{TT}(\theta)$ at large angles.

The way that the cross correlation of the temperature and lensing fields will be affected depends on 
 the underlying details contained in the metric potential.  This means that predictions of how $C^{T\varphi}$ will be affected by new physics can only occur after choosing a particular model.
Consequently, absent a specific alternative model we cannot construct a statistic that fits
 well into a Bayesian statistical approach and differentiates between the fluke hypothesis and all other models generically.
The approach we therefore take is to use $\Lambda$CDM (without the constraints imposed by the fluke hypothesis) as the comparison model
for its predictions of the statistical properties of $C^{T\varphi}$ and then construct as $S_{1/2}^{T\varphi}$-like
 statistic that optimizes
the ability to select between the fluke hypothesis and this comparison model.  
The statistic we propose can thus be used to falsify the prediction of the fluke hypothesis.

%This is how we will test the null hypothesis that our universe is a statistical anomaly. We will assume that the 
%$C^{T\varphi}(\theta)$ spectrum follows from the standard $\Lambda$CDM prediction, but include the knowledge
%that our temperature field has the observed features. We will then calculate statistics from this $C^{T\varphi}(\theta)$.
%Future observations which fell nicely within the distribution of these statistics or in the tails of our statistic 
%would either lend evidence to or rule out the null hypothesis respectively.

%%%%%%%%%%%%%%%%%%%%%%%%%%%%%%%%%%%
\section{Constrained Sky Realizations}\label{constrained}
%%%%%%%%%%%%%%%%%%%%%%%%%%%%%%%%%%%

We know that  we live in a Universe with a particular angular power spectrum of the temperature, $C^{TT}_\ell$, 
and a particular value of $S^{TT}_{1/2}$, and we must see what including this as a prior constraint on the allowed
realizations of $\Lambda$CDM does to the probability distribution of values of our target statistic.

It is well known how to create realizations of ordinary $\Lambda$CDM with a fixed set of parameters.
 It is more unusual to create constrained 
realizations of $\Lambda$CDM -- ones that reproduce, within the measurement errors, the angular power spectrum
of the observed sky, and with both a full sky and a cut sky $S^{TT}_{1/2}$ no larger than those of the observed sky. 
A detailed description of how create constrained realizations is contained in section 2 of \citep{Copi:2013zja}.
Briefly, we treat the observational errors in the {\it WMAP}-reported $C^{TT}_\ell$ 
as Gaussian distributed and generate many random $C^{TT}_\ell$ 
from Gaussian distributions centred on the {\it WMAP}-reported values, and correct
this realization for the slight correlation induced on partial skies.  
The resulting sky realization is guaranteed to have a full-sky $S_{1/2}^{TT}$ consistent with the small value
in the full-sky {\it WMAP} ILC map. We only keep realizations with an $S_{1/2}^{TT}$ less than 
the observed cut-sky calculated value ($1292.6\;\umu \rmn{K}^4$ for {\it WMAP}-7 and $1304\;\umu \rmn{K}^4$ for {\it WMAP}-9) for analysis.

With a set of $10^{5}$ such constrained temperature realizations, $a_{\ell m}^T$, 
we can compute the corresponding set of constrained lensing potential realizations, $a_{\ell m}^{\varphi}$.
This is done using standard techniques for generating correlated random variables using
 the {\small HEALPix} package\footnote{ http://healpix.sourceforge.net}\citep{Gorski:2004by} and is reviewed
in the appendix of \citep{Copi:2013zja}. The required 
 input spectra $C_{\ell}^{T\varphi}$ and $C_{\ell}^{\varphi\varphi}$ were generated 
 with \small{CAMB}\footnote{http://camb.info}\citep{Lewis:1999bs}.

A full-sky analysis of the CMB relies on a reconstruction of the fluctuations behind the Galactic cut.
we instead use the cut-sky measured value as a threshold to avoid any bias that the reconstruction might induce. We emphasize that no attempt is made (nor is any necessary) to argue that the cut-sky statistic is a 
better estimator of the value of some full-sky version of the statistic on the  full-sky (if we could measure it reliably) 
or on the ensemble.   The cut-sky statistic need only be taken at face value as a precise prescription for something
that can be calculated from the observable sky.  Calculating $S_{1/2}$ 
with partial sky data has a well defined procedure --  statistics are just calculated using 
pseudo-$C_{\ell}$s without reference to any  full sky estimators.
A more detailed discussion of this can be found in~\citep{Copi:2013zja}.

%%%%%%%%%%%%%%%%%%%%%%%
\section{calculating $S_{1/2}$ and $S_{x}$ statistics}\label{s12}
%%%%%%%%%%%%%%%%%%%%%%%%

Once we have a set of $a_{\ell m}^T$ and $a_{\ell m}^{\varphi}$, we can calculate $S_{1/2}^{T\varphi}$ as above (\ref{s12eq}). However, instead of using  $C(\theta)$ directly, we calculate the statistic 
using $C_{\ell}$s:
\begin{equation}
\label{eq:Shalf}
  S_{1/2}^{XY} \equiv \int_{-1}^{1/2}\dderiv(\cos\theta) [C^{XY}(\theta)]^2 
  = \sum_{\ell=2}^{\ell_{\mathrm max}} C_{\ell}^{XY} {I}_{\ell\ell'} C_{\ell'}^{XY}.
\end{equation}
 In computing Eq.~\ref{eq:Shalf} we used an $\ell_{\rmn{max}}=100$, as the $C_{\ell}$ fall sharply and higher order 
 modes have a negligable contribution to the statistic. An explicit definition of the 
 $I_{\ell \ell^{\prime}}$ matrix 
 can be found in Appendix B of \citep{Copi:2008hw}.

For temperature-lensing cross correlations, we can
optimize the statistic
{\it a priori} so that it best discriminates between
constrained realizations of $\Lambda$CDM and unconstrained $\Lambda$CDM. To do this, we generalize
$S_{1/2}$ to 
\begin{equation}\label{sxyeq}
S_{a,b}^{XY}=\int_{a}^{b}\mbox{d}(\cos\theta)[C^{XY}(\theta)]^2  \quad \mbox{for} \quad -1\leq a< b\leq1.
\end{equation}

For each possible pair of $a=\cos(\theta_{a})$ and $b=\cos(\theta_{b})$, we calculate the distribution of 
$S_{a,b}^{T\varphi}$ values 
for ensembles of both constrained and unconstrained realizations.
We compute the $99$-per centile value for the constrained distribution (i.e. the value of $S_{a,b}^{T\varphi}$ that is
 greater than $99$ per cent of the members of the constrained ensemble).    We then determine the fraction
of the values in the $S_{a,b}$ distribution for
 the unconstrained ensemble that are {\it larger} than the $99$ per cent constrained value.
   The higher the percentile,
the better $S_{a,b}^{T\varphi}$ is at discriminating between the constrained and unconstrained models.     We repeat
the analysis for $99.9$-percentile. We choose two different confidence levels for analysis here to show that 
optimization will lead to different $(a,b)$ ranges. The confidence level and corresponding per centage for 
optimization should always be chosen before 
any analysis on a particular data set is carried out. This process was performed using reported $C_{\ell}^{TT}$ and corresponding error bars from both 
{\it WMAP} 7- and {\it WMAP} 9-year releases.\footnote{http://lambda.gsfc.nasa.gov/}.

%%%%%%%%%%%%%%%%%%%%%%%%%
\section{Results}\label{results}
%%%%%%%%%%%%%%%%%%%%%%%%

\begin{figure}
\includegraphics[scale=.6]{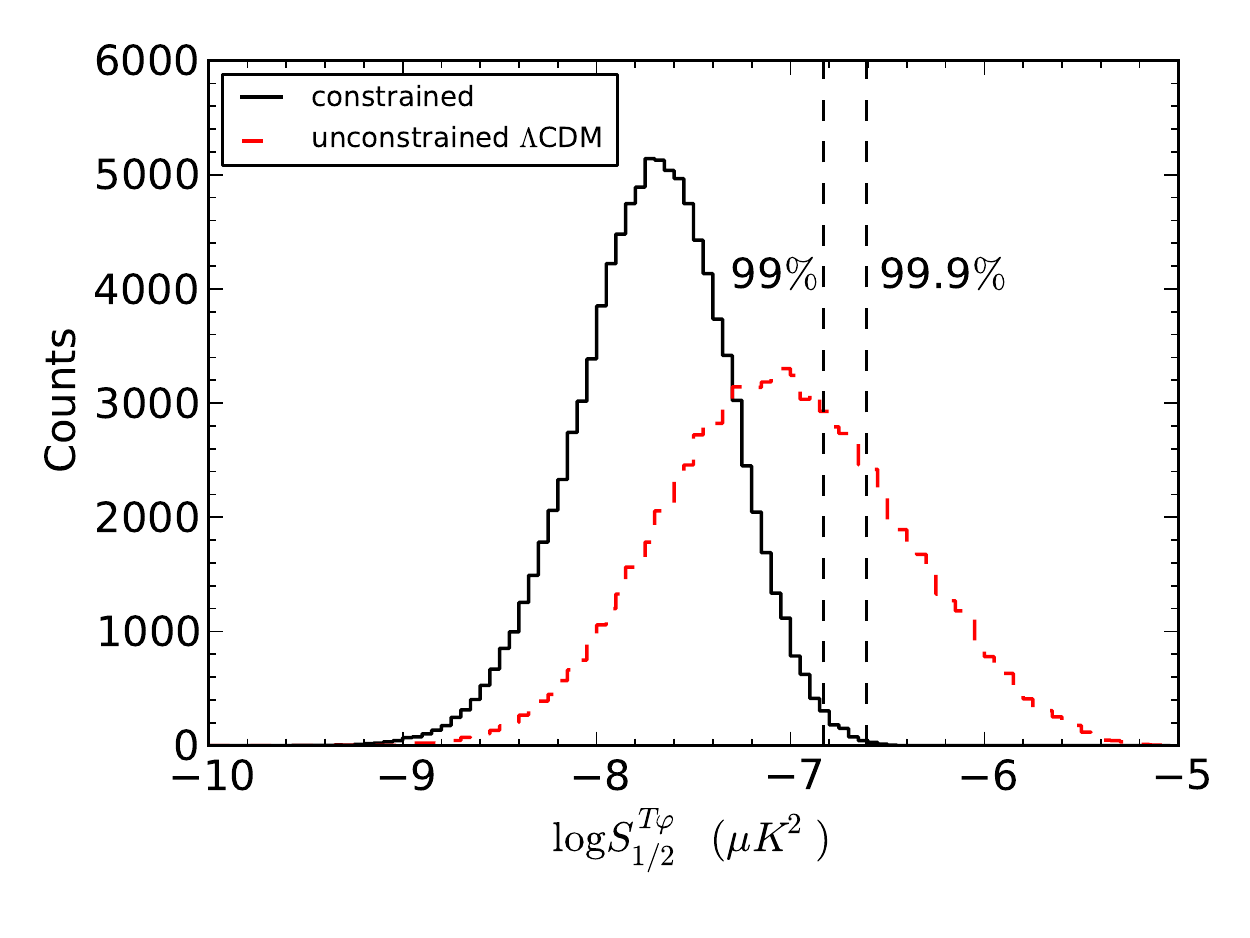}
\caption{Histogram of $S_{1/2}^{T\varphi}$ values for $10^5$ simulations for constrained (black solid) and
unconstrained (red dashed) $\Lambda$CDM realizations from {\it WMAP} 9 parameters. The dashed lines show  the $99$-per centile and $99.9$-per centile 
values for the constrained realizations.}\label{s12fig9yr}
\end{figure}

The distributions on a cut sky for unconstrained $\Lambda$CDM and for $\Lambda$CDM constrained by the {\it WMAP} 9-year power spectrum and $S_{1/2}^{TT}$ value are shown in Fig.~\ref{s12fig9yr}. 
The two distributions have a significant overlap. The analysis shows
$39.6$ per cent of the 
unconstrained $\Lambda$CDM values fall above the $99$-per centile constrained value of $1.48\times 10^{-7}\;\umu\rmn{K}^2$, 
and $26.7$ per cent of the unconstrained $\Lambda$CDM values falling above the $99.9$-per centile constrained value of $2.47\times 10^{-7}\;\umu\rmn{K}^2$.

 We repeated this analysis for the {\it WMAP} 7-year release, and found
a negligible difference between results generated with the 7- and 9-year data sets. 
We conclude that the changes in {\it WMAP} reported values and error bars and the 
best-fitting model for $C_{\ell}^{TT}$  by release year do
not affect the results in any significant way. We have also calculated all of the generalized 
$S_{a,b}^{T\varphi}$ statistics for
the 7- and 9-year data, and found that it as well produced similar results. Therefore, we will proceed with presenting only the results from the {\it WMAP} 9-year analysis.

%\begin{table}
%\begin{tabular}{| r || p{1cm} | p{2.5cm} | p{1cm} | p{2.5cm} | }
%\hline
%       & $99\%$ value & $\% >$ unconst. $\Lambda$CDM & $99.9\%$ value & $\% >$ unconst. $\Lambda$CDM \\ \hline
%7-year &   &   &   \\ \hline
%9-year &   &   &   \\ \hline
%\end{tabular}
%\caption{values}
%\end{table}

\begin{figure}
\includegraphics[scale=.5]{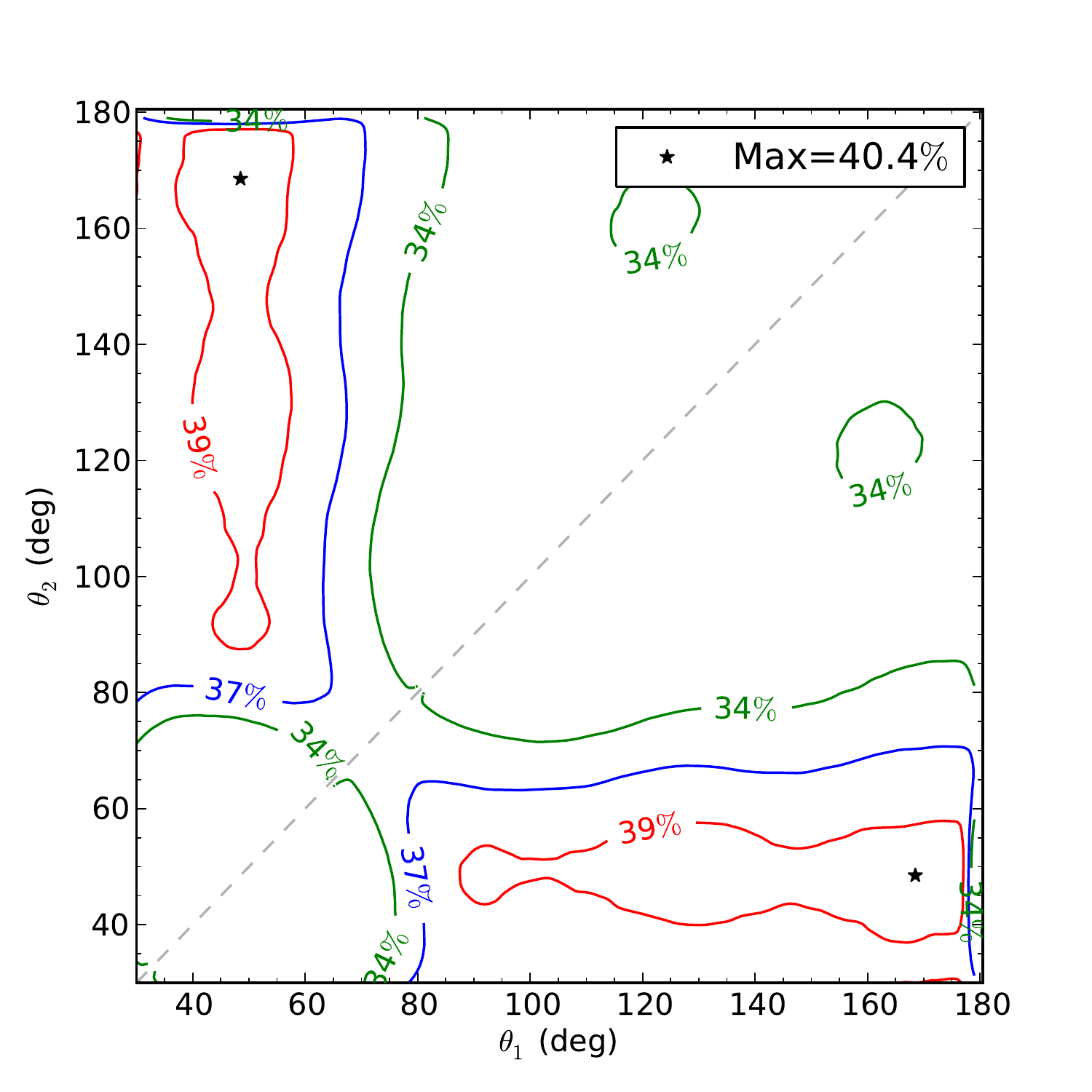}
\caption{Contour plot showing percentage of unconstrained $\Lambda$CDM $S_{a,b}^{T\varphi}$ values falling above
the $99$-per centile value for the constrained realizations. Stars show the maximum difference, which occurs for 
an integral over the range of 48$\degr$ to 168$\degr$.}\label{sxy99}
\end{figure}

\begin{figure}
\includegraphics[scale=.5]{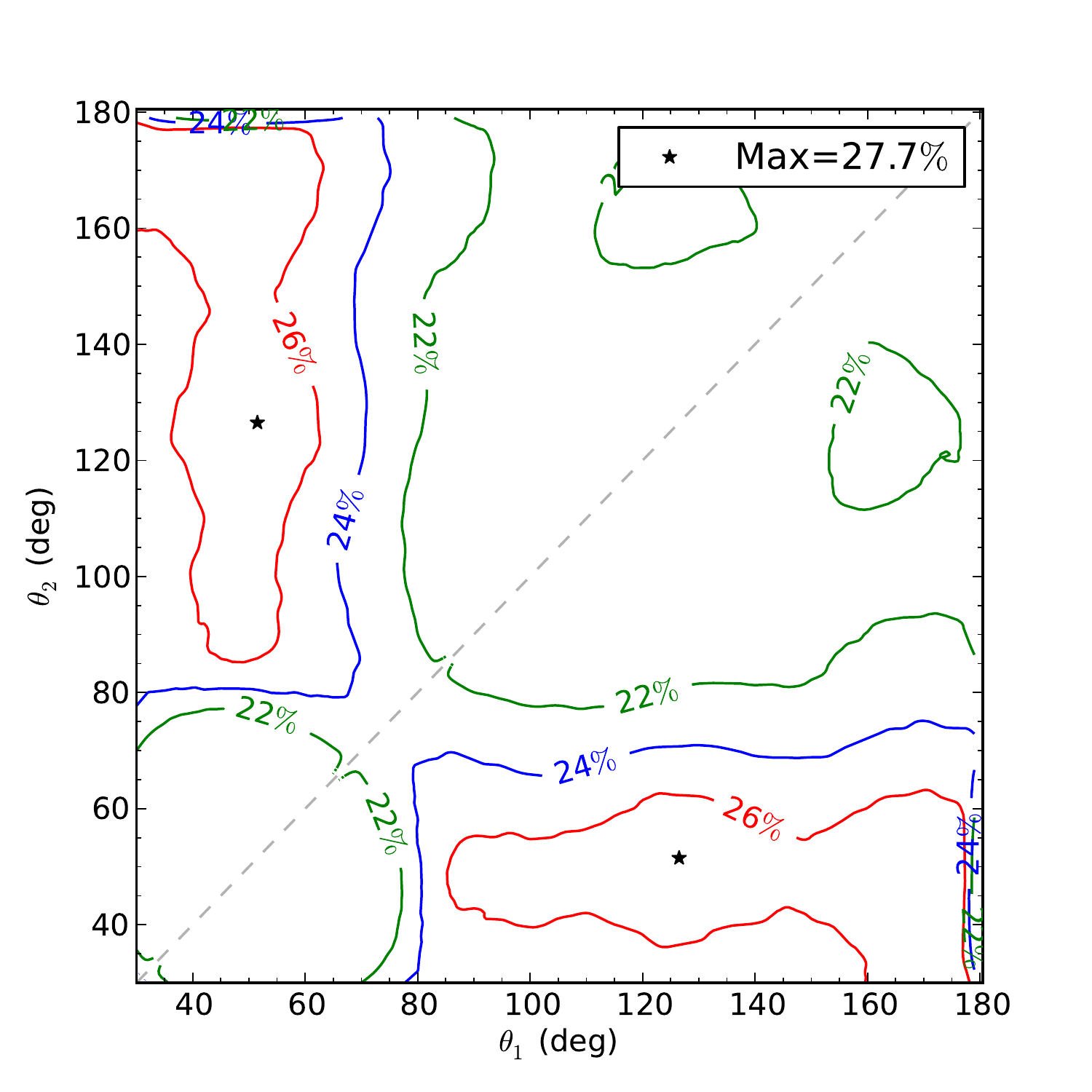}
\caption{Contour plot showing percentage of unconstrained $\Lambda$CDM $S_{a,b}^{T\varphi}$ values falling above
the $99.9$-per centile value for the constrained realizations. Stars show the maximum difference, which occurs for 
an integral over the range of 53$\degr$ to 127$\degr$.}\label{sxy999}
\end{figure}

\begin{figure}
\includegraphics[scale=.6]{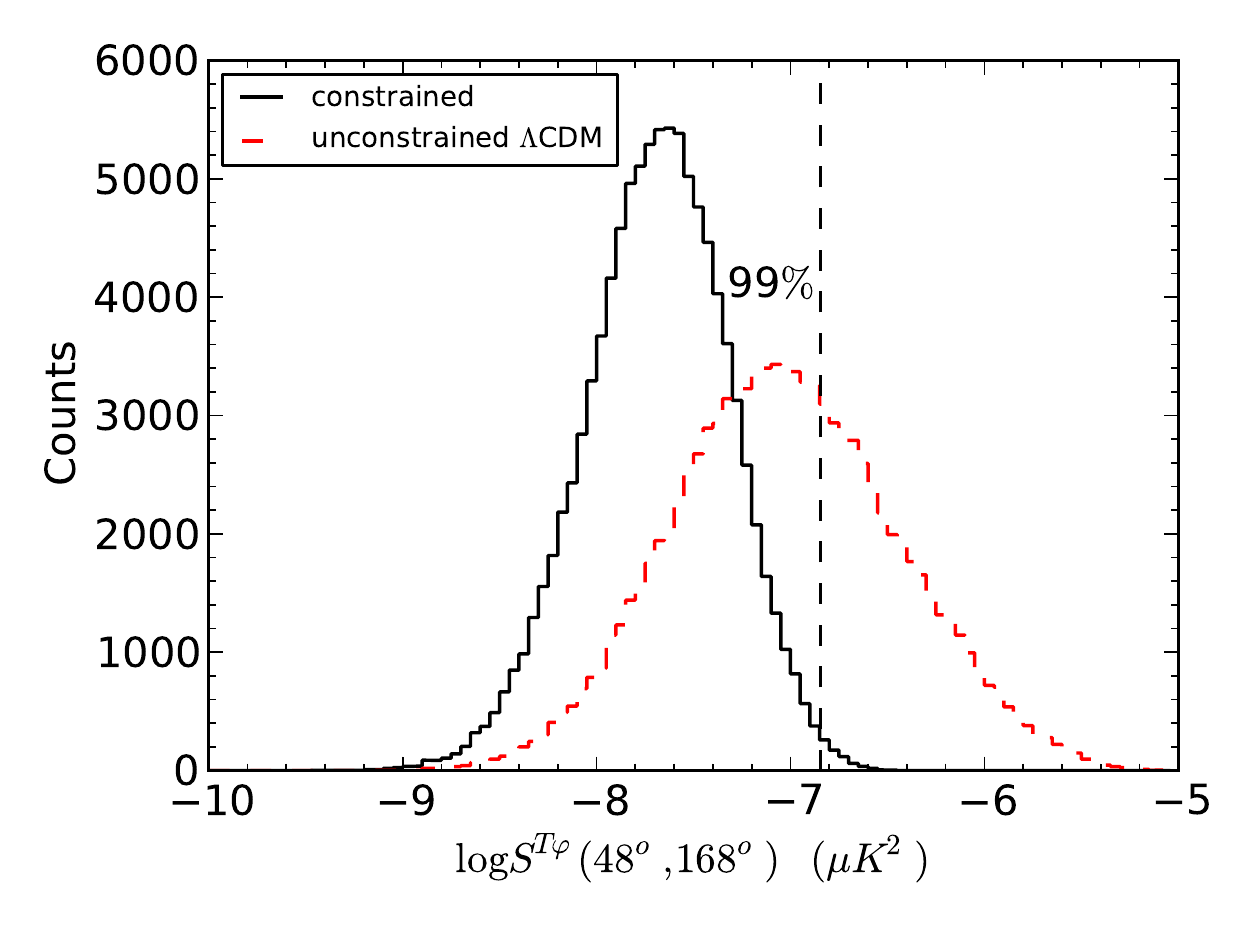}
\caption{Histogram showing distribution of constrained (black solid) and unconstrained (red dashed) $\Lambda$CDM $S_{a,b}^{T\varphi}$ values for 
$\theta_a=168\degr, \theta_b=48\degr$. This range of angles gives the most optimal statistic for ruling out the null
hypothesis at the $99$ per cent level.}\label{hist99}
\end{figure}

\begin{figure}
\includegraphics[scale=.6]{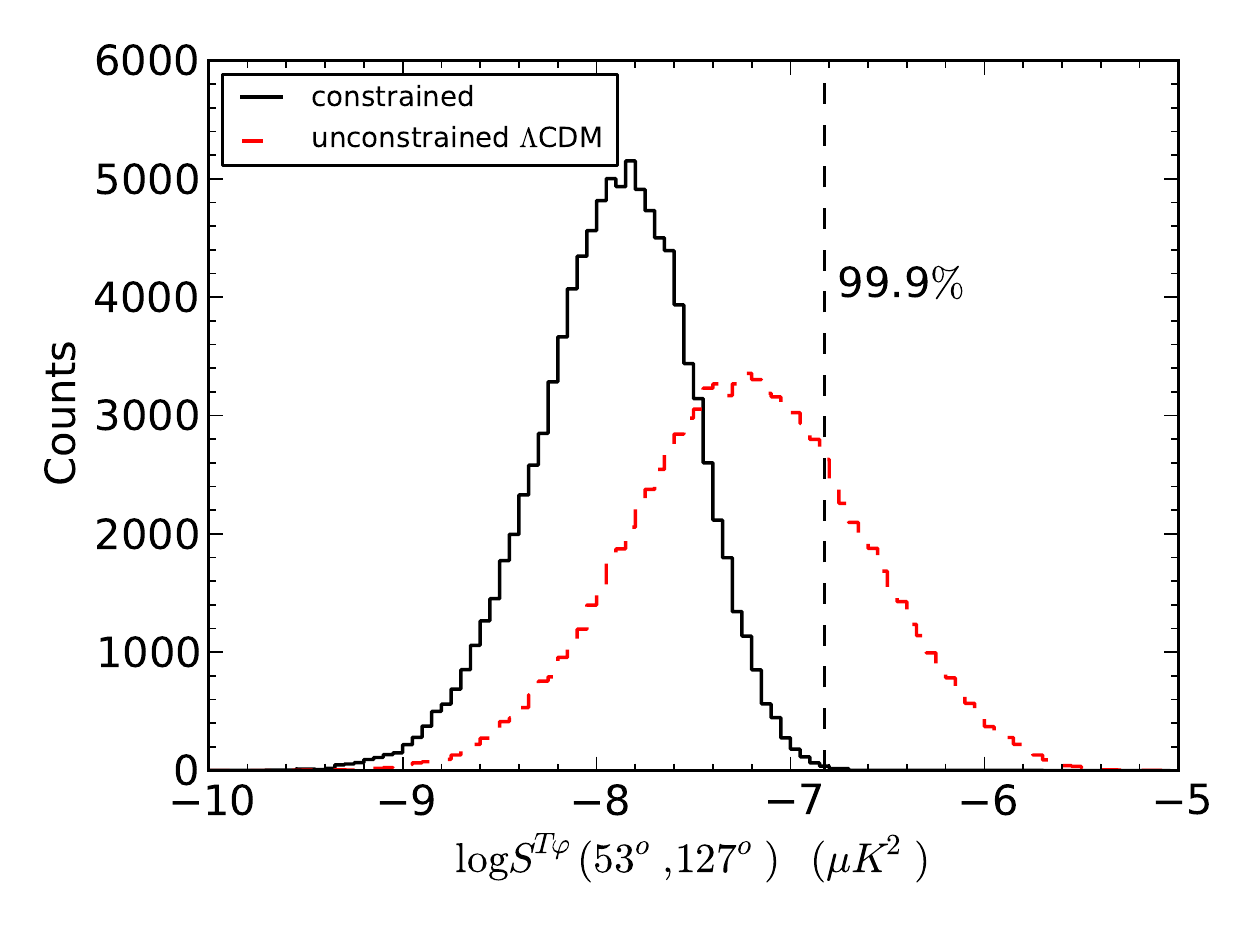}
\caption{Histogram showing distribution of constrained (black solid) and unconstrained (red dashed) $\Lambda$CDM $S_{a,b}^{T\varphi}$ values for 
$\theta_a=127\degr, \theta_2=53\degr$. This range of angles gives the most optimal statistic for ruling out the null
hypothesis at the $99.9$ per cent level.}\label{hist999}
\end{figure}

Figs.~\ref{sxy99} and~\ref{sxy999} show only a  small improvement by choosing to integrate over a range of angles
other than $60\degr$ to $180\degr$.   The maximum discriminating power is $40.4$ per cent for the statistic  integrating over the  range
$ a=\cos(168\degr),b=\cos(48\degr) $.  For the $99.9$-per centile, the maximum discriminating power  
drops to $27.8$ per cent, with  the optimal range of angles changing
slightly to  $a=\cos(127\degr),b=\cos(53\degr) $. Figs.~\ref{hist99} and ~\ref{hist999} show the distribution of 
the $S_{a,b}$ statistic from Eq.~\ref{sxyeq} for these ranges of angle-cosines. The $99$-per centile value of the statistic for $\theta_a =168\degr, \theta_b =48\degr$ on 
the ensemble of constrained realizations is $1.43\times10^{-7}\;\umu\rmn{K}^2$, and 
the $99.9$-per centile value for $\theta_a =127\degr, \theta_b =53\degr$ is 
$1.50\times10^{-7}\;\umu\rmn{K}^2$.

We have not repeated the analysis for the first {\it Planck} temperature maps because an approximate covariance matrix is not yet available, but we expect the results to be similar based on the close match between large-scale features in the 
{\it WMAP} and {\it Planck} sky maps.

%\begin{figure}
%\includegraphics[scale=.6]{S12_hist_7yr_newax.pdf}
%\caption{Histogram of $S_{1/2}^{T\varphi}$ values for $10^5$ simulations for constrained (black solid) and
%unconstrained (red dashed) $\Lambda$CDM realizations from {\it WMAP} 7 parameters. The dashed lines show the $99$-percentile %and $99.9$-percentile 
%values for the constrained realizations.}\label{s12fig}
%\end{figure}

%%%%%%%%%%%%%%%%%%%%%%%%
\section{Conclusions}\label{conclusions}
%%%%%%%%%%%%%%%%%%%%%%%%

In light of the new data from {\it Planck}, large-angle anomalies have been gaining more traction in the community
as potential evidence of interesting primordial physics that deviates from the widely 
accepted $\Lambda$CDM paradigm. 
In particular, the observed lack of temperature-temperature correlations at angles larger than $60\degr$  
has been characterized by the $S_{1/2}^{TT}$ statistic, and the measured value has shown to occur in less than $.1$ per cent 
 of standard $\Lambda$CDM realizations on a 9-year KQ85 masked sky~\citep{Copi2004, Copi:2013cya}. 
 Since $S_{1/2}^{TT}$ was an {\it a posteriori} choice of a statistic after seeing the shape of the two-point function from the {\it WMAP} data, 
the possibility that our Universe happens to just be a statistical fluke has been advocated. We wish to test
this  hypothesis by calculating statistics for a different correlation function whose distribution would be expected to 
be markedly altered (compared to normal) in realizations of $\Lambda$CDM that reproduce the observed small $S^{TT}_{1/2}$.
For this purpose we have used the cross-correlation between
 temperature and lensing potential $\varphi$ as an {\it a priori} test of the null hypothesis.

We have calculated the distribution of the $S^{T\varphi}_{1/2}$ statistic for $10^5$ realizations of unconstrained $\Lambda$CDM 
with the {\it WMAP} 7 and {\it WMAP} 9 best-fitting cosmological parameters,  as well as a similar number of 
realizations constrained to have a value for $S_{1/2}^{TT}$ no larger than the observed value, and found no 
significant difference between results calculated from each data set
despite changes to the reported central values and error bars. We showed that  $39.2$ per cent of members of the ensemble of unconstrained realizations had
$S^{T\varphi}_{1/2}$ greater than the $99$-per centile value of $S^{T\varphi}_{1/2}$ in the ensemble of constrained realizations.
This represents a modest (but not insignificant) ability to distinguish strongly between the constrained and unconstrained models.

We have also defined a generalised statistic for testing the null hypothesis by investigating which pair
of angles used in calculating $S_{a,b}^{T\varphi}$ defined in Eq.~\ref{sxyeq} provided the largest per centage of 
unconstrained $\Lambda$CDM $S_{a,b}^{T\varphi}$ lying above the $99$-per centile value for the constrained distribution. 
We find that restricting the integration range from $48\degr$ to $168\degr$ slightly improves the ability of the statistic to rule 
out a given measured value being consistent with constrained $\Lambda$CDM. 
To rule out the fluke hypothesis at the $99.9$-per centile level, we found that the optimal range of angles is $53\degr$ to $130\degr$. 
The contours showing the descriminating power of the $S^{T\varphi}_{a,b}$ statistic for the constrained $\Lambda$CDM versus
unconstrained $\Lambda$CDM were shown in Figs.~\ref{sxy99} and~\ref{sxy999}, and the histograms for the $S_{a,b}^{T\varphi}$
statistic for the optimal angular ranges were shown in Figs.~\ref{hist99} and~\ref{hist999}.   
However, because the improvement over the $S^{T\varphi}_{1/2}$ statistic is modest, and because $S^{T\varphi}_{a,b}$ is
optimized to select between constrained $\Lambda$CDM and unconstrained $\Lambda$CDM,
simplicity argues for using $S^{T\varphi}_{1/2}$ to parallel previous analysis of $C_{\ell}^{TT}$.

%It should be noted, though, that because the histograms of the unconstrained and constrained $\Lambda$CDM realizations
%overlap with one other, the fluke hypothesis could be ruled out at a particular confidence level, but {\it not} be in
%contention with the generic $\Lambda$CDM cosmology. This would mean that the hypothesis that our universe is a statistical
%fluke could be ruled out while also agreeing with $\Lambda$CDM predictions. There could be several (not definitive) 
%explanations for this type of result, such as an error in the analysis of our temperature data at large
%angles, something else, or something else.

The outlined procedure for assessing the discriminating power
of the $T\varphi$ cross-correlation statistics can be used as a generic prescription for optimizing statistics of any cosmological data. Choice of a particular confidence level to optimize for should always be made before calculations of any 
$S_{1/2}$-like statistics are carried out to avoid any bias in reporting results.   

Clearly, some model comparisons will provide statistics from $T\varphi$ that are more useful than others. 
In particular, since the $\varphi$ field
is dominated by effects inside the last-scattering surface, it has a large correlation with $\Theta_{\rmn{ISW}}$. This
means that unless a proposed model can find some way to supress this particular term, there will not be a sharp difference
for $S_{a,b}^{T\varphi}$ statistics from $\Lambda$CDM, which will limit its usefulness if one
prefers to compare their model to $\Lambda$CDM. Regardless, $T\varphi$ correlations will provide an important consistency check for the data, as the
large $\Theta_{\rmn{ISW}}\varphi$ contribution is a probe of physics on the interior of our Hubble volume. It
 is therefore complimentary to the $TT$ signal which is largely comprised of effects at the last-scattering surface.
 
The $S^{T\varphi}$ statistic is not particularly helpful for testing a hypothesis that the underlying gravitational potential fluctuations lack correlations on scales larger than some comoving scale subtending $60\degr$ at the redshift of last scattering.
Any suppression in $\langle\Phi(\boldsymbol{  \hat{n}_1})\Phi(\boldsymbol{  \hat{n}_2})\rangle$ that gives
rise to the observed $TT$ spectrum would {\it not} have any significant effect on the shape of $C^{T\varphi}(\theta)$
compared to $\Lambda$CDM. This fact is precisely due to the large $\Theta_{\rmn{ISW}}\varphi$ term.
 
Other cross correlations may prove to be more fruitful. For example, 21 centimeter emission correlated with 
temperature fluctuations will partially mitigate the large $\Theta_{\rmn{ISW}}\varphi$ problem. The 21 centimeter 
emission spectrum comes to us from localized region of redshift space, and does not have a component which is an integral 
along the line of sight. This in particular should reduce the magnitude of the correlation compared to $C^{T\varphi}(\theta)$.
In a future work, we will show how viable this cross correlation will be for testing the null hypothesis, as well as
provide predictions for the shape of $C(\theta)^{T\;21\rmn{cm}}$ with an imposed cutoff.

A related calculation \citep{Copi:2013zja} examined the implications of the fluke hypothesis for the temperature-polarization angular correlation function,
in particular the correlation of temperature with the Stokes $Q$ parameter, $C^{TQ}(\theta)$. This statistic excludes 
the fluke hypothesis at 99.9 per cent C.L. for 26 per cent of realizations of unconstrained $\Lambda$CDM, or at 99 per 
cent C.L. for 39 per cent of such realizations.

In summary, the temperature auto-correlation of the CMB sky behaves oddly at large angular separations.
No satisfactory current theory explains this anomaly, and so the leading explanation is that it is a statistical fluke.
In the absence of specific models to compare directly with $\Lambda$CDM, the best strategy is to identify other measurable quantities 
with probability  distributions that are affected by the knowledge that $S^{TT}_{1/2}$ is small, and make predictions for 
the new probability distribution functions.  In this way we can test the fluke hypothesis with
 current and near-future CMB data sets.

%This work provides {\it a priori} statistics and predictions that can be used for testing the hypothesis that 
%the two-point function of the CMB tempreature that we observe is a statistical fluke. Because we know that we have
%a universe where the temperature data has a lack of power at large angles, we expect that statistics calculated from
%the cross-correlation of temperature and lensing potential observations {\it must} lie within the constrained distribution
%for that statistic. When a value of $S_{1/2}$ or the more optimized $S_{a,b}$ is calculated from observations, we will be able
%to rule out (or lend weight to) the hypothesis that we live in a universe that is a statistical fluke 
%at the appropriate confidence level.

%%%%%%%%%%%%%%%%%%%%%%%%%%%%
\section*{acknowledgements}
%%%%%%%%%%%%%%%%%%%%%%%%%%%%

The authors would like to thank Simone Aiola for useful discussions.The numerical simulations were performed 
on the facilities provided by the Case ITS High Performance Computing Cluster.
AY is supported by NASA NESSF Fellowship.  CJC, GDS and AY are supported
by a grant from the US DOE to the Particle Astrophysics Theory group at CWRU. 
AK has been partly supported by NSF grant AST-1108790.

\end{document}